\documentclass[12pt,a4paper]{article}
\pdfoutput=1
\usepackage{a4wide}
\usepackage[centertags]{amsmath}
\usepackage{amssymb}
\usepackage{cite}
\usepackage{ulem}
\usepackage{float}
\usepackage[pdftex]{graphicx,color} 
\allowdisplaybreaks
\begin{document}
\thispagestyle{empty}
\newenvironment{gleichung}{\begin{equation}\begin{aligned}}{\end{aligned}\end{equation}\noindent}
\newenvironment{gleichung*}{\begin{equation*}\begin{aligned}}{\end{aligned}\end{equation*}\noindent}

\begin{center}
{\Large  \bf Towards the Holographic Dual of $\mathbf{{\mathcal N}=2}$ SYK} 
\end{center}

\vspace*{1cm}

\centerline{Stefan F{\"o}rste, Joshua Kames-King and Max Wiesner}
\vspace{1cm}

\begin{center}{\it
Bethe Center for Theoretical Physics\\
{\footnotesize and}\\
Physikalisches Institut der Universit\"at Bonn,\\
Nussallee 12, 53115 Bonn, Germany}
\end{center}

\vspace*{1cm}

\centerline{\bf Abstract}
\vskip .3cm
The gravitational part of the holographic dual to the SYK model has
been conjectured to be Jackiw-Teitelboim (JT) gravity.
In this paper we construct an AdS$_2$ background in $\mathcal{N}=(2,2)$ JT gravity and show that the gravitational dynamics are -- as in the $\mathcal{N}=0 $ and ${\mathcal N}=1$ cases -- fully captured by the extrinsic curvature as an effective boundary action. This boundary term is given by the  super-Schwarzian of the $\mathcal{N}=2$  SYK model, thereby providing further evidence of the JT/SYK duality. The chirality of this SYK model is reproduced by the inherent chirality of axial $\mathcal{N}=(2,2)$ supergravity.

\vskip .3cm

\newpage
\section{Introduction}

The Sachdev-Ye-Kitaev (SYK) model \cite{Sachdev:1992fk,Kitaev,Sachdev:2010um,
Sachdev:2015efa,Polchinski:2016xgd,Jevicki:2016ito,Maldacena:2016hyu} is 
conformally (i.e.\ reparametrisation) invariant in the IR. 
The breaking of conformal symmetry results in an effective Lagrangian
for time reparametrisations which is given by the Schwarzian.
Models without random couplings sharing this property have been
constructed and studied in e.g.\
\cite{Gurau:2016lzk,Witten:2016iux,Klebanov:2016xxf,
Krishnan:2016bvg,Krishnan:2017ztz,Peng:2017kro,
Narayan:2017qtw,Krishnan:2017txw,Choudhury:2017tax,Krishnan:2017lra,
BenGeloun:2017jbi}. 
There are various other mutations of the SYK model, for instance,
higher dimensional analogs have been proposed in
\cite{Davison:2016ngz,Berkooz:2017efq,Turiaci:2017zwd}, 
complex versions are studied
e.g.\ in
\cite{Davison:2016ngz,Bulycheva:2017uqj,Bhattacharya:2017vaz},
more 
than one flavour and non Abelian global symmetries have been
investigated in \cite{Gross:2016kjj,Yoon:2017nig,Chen:2017dbb,
Zhang:2017jvh}. For
the present 
paper the supersymmetric versions constructed in
\cite{Fu:2016vas} are most relevant. For variations and further
aspects of
supersymmetric SYK models see
\cite{Nishinaka:2016nxg,Sannomiya:2016mnj,Peng:2016mxj,Li:2017hdt,
Stanford:2017thb, 
Mertens:2017mtv,Kanazawa:2017dpd,Murugan:2017eto,
Yoon:2017gut,Peng:2017spg,Hunter-Jones:2017raw,Hunter-Jones:2017crg,
Narayan:2017hvh}.  
Now, superreparametrisations are an exact symmetry only in the
infrared limit, and their breaking gives rise to an effective 
super-Schwarzian action. 

The holographic dual is believed to contain some version of dilatonic 2d 
gravity arising quite universally in compactifications from higher dimensions
\cite{Almheiri:2014cka}. Moreover, near the horizon of an AdS$_2$ black hole, 
corresponding to  
the IR of SYK, solutions of the dilaton are nearly constant. 
Approximating to linear order in non constant contributions leads to Jackiw-Teitelboim (JT) gravity \cite{Jackiw:1984je,Teitelboim:1983ux}, 
for a recent review see \cite{Sarosi:2017ykf}. JT gravity has been 
corroborated as the gravitational dual of the SYK model e.g.\ by deriving 
the Schwarzian as an effective action for the UV regulator curve 
\cite{Turiaci:2016cvo,Jensen:2016pah,Maldacena:2016upp,Cvetic:2016eiv}.
Liouville theory instead of JT gravity has been considered in
\cite{Mandal:2017thl}.     
A three dimensional holographic dual has, however, been advocated in a series 
of papers \cite{Das:2017pif,Das:2017hrt,Das:2017wae}. Corrections to JT gravity 
have been recently proposed in \cite{Kitaev:2017awl}. Gathering information 
about the holographic dual beyond the gravitational sector has been the 
subject of \cite{Gross:2017hcz,Gross:2017aos}.

To start extending these investigations to supersymmetric versions of SYK it 
was shown in \cite{Forste:2017kwy} that an ${\mathcal N}=(1,1)$
supersymmetric version of  
JT gravity \cite{Chamseddine:1991fg} supplemented with the appropriate 
boundary term leads to the ${\mathcal N}=1$ super-Schwarzian as an
effective action for the UV regulator curve. In the present paper we
will extend this further to ${\mathcal N} =(2,2)$ JT gravity. 

The paper is organised as follows. In section 2 we collect some results
on ${\mathcal N}=(2,2)$ supergravity in superconformal gauge. Section
3 deals with the ${\mathcal N} =(2,2)$ extension of JT gravity. The
Gibbons-Hawking-York term is added. In section 4 the superconformal gauge is solved for AdS$_2$ as a supersymmetric background. The main result, the super-Schwarzian as effective Lagrangian of the boundary curve, is obtained in section 5. In section 6, a consistency check will be
performed. Our results are summarised in section 7.  
\section{$\mathbf{{\mathcal N}=(2,2)}$ Supergravity}

In this section, we collect some information about extended ${\mathcal
  N}=(2,2)$
supersymmetry and supergravity in two dimensions. Useful references
are \cite{Howe:1987ba,West:1990tg,Grisaru:1994dm,Grisaru:1995dr,
Closset:2014pda,Gomis:2015yaa}. The two dimensional rigid
${\mathcal N}=(2,2)$ superspace is given by the coset space
\cite{West:1990tg} 

\begin{align*}
    \frac{(2,2) \text{Supergroup}}{\text{Lorentz}\otimes U_A(1)
  \otimes U_V(1)}\,, 
\end{align*}
with coordinates 
\begin{align*}
    z^\pi=(z,\theta^+,\bar \theta^+;\bar z, \theta^-, \bar \theta^-)\,
\end{align*}
and covariant derivatives 
\begin{align*}
    \partial_z\,,\;\; D_+=\frac{\partial}{\partial
  \theta^+}+\frac{1}{2}\bar \theta^+ \partial_z\,,\; \;\bar
  D_+=\frac{\partial}{\partial \bar \theta^+}+\frac{1}{2}
  \theta^+ \partial_z , \\  
     \partial_{\bar z}\,,\;\; D_-=\frac{\partial}{\partial
  \theta^-}+\frac{1}{2}\bar \theta^- \partial_{\bar z}\,,\;\; \bar
  D_-=\frac{\partial}{\partial \bar \theta^-}+\frac{1}{2}
  \theta^-\partial_{\bar z} .\;  
\end{align*}
They satisfy the anticommutation relations 
\begin{align*}
    \{D_+,\bar D_+\}=\partial_z\,,\;\;\; \{D_-,\bar D_-\}=\partial_{\bar z}\,.
\end{align*}
There are two versions of minimal $\mathcal{N}=2$ supergravity, which
can be obtained from the nonminimal $U_A(1)\otimes U_V(1)$ by gauging
either the $U_A(1)$ or $U_V(1)$ factor of the tangent space symmetry
group \cite{Grisaru:1994dm}. Here we will focus on the axial version
of minimal $\mathcal{N}=2$ sugra with gauged $U_A(1)$, which can
also be obtained by dimensionally reducing $\mathcal{N}=1$ sugra in
$d=4$. Accordingly, the tangent space symmetry group consists of the
2D Lorentz group and the gauged $U_A(1)$ factor. 

The spinorial covariant derivatives in the minimal theory are given by 
\begin{equation}
    \nabla_\alpha=E_\alpha + \Omega_\alpha \mathcal{J} + \Sigma_\alpha
  \mathcal{Y}\,,\label{eq::Nabla1} 
\end{equation}
where $\alpha=\pm$ is a flat space spinor index and  $\mathcal{J},
\,\mathcal{Y}$ are respectively 
the Lorentz and $U_A(1)$ generators with corresponding connections
$\Omega_\alpha$ and $\Sigma_\alpha$. For the complex conjugates and
vector derivatives similar relations hold.  

The Lorentz and $U_A(1)$ generators form together with the four
supercharges $Q^+$, $Q^-$, $\bar Q^+$, $\bar Q^-$ our SUSY algebra~\cite{West:1990tg}: 
\begin{gleichung}
\left[Q_+,\mathcal{Y}\right]&=-Q_+\,, &\left[\bar Q_+,\mathcal{Y}\right]&=
\bar Q_+\, ,\\
\left[Q_-,\mathcal{Y}\right]&=Q_-\,, & \left[\bar Q_-,\mathcal{Y}\right]&=
-\bar Q_-\, ,\\
\left[Q_+,\mathcal{J}\right]&=\frac{i}{2}Q_+\,,& \left[\bar Q_+,\mathcal{J}
\right]&=\frac{i}{2}\bar Q_+\, ,\\
\left[Q_-,\mathcal{J}\right]&=-\frac{i}{2}Q_-\,, &\left[\bar Q_-,\mathcal{J}
\right]&=-\frac{i}{2}\bar Q_- \, .
\end{gleichung}
Therefore Majorana constraints are implemented by the following constraint on Weyl spinors in Euclidean space:
\begin{align}
 (Q_+)^{*}=   \bar Q_- \,  \, , \;\;\;\;  (Q_-)^{*}=   \bar Q_+ \, .
\end{align}
For convenience, we introduce the following linear combination of the
tangent group generators:  
\begin{align}
    M&\equiv \mathcal{J}-\frac{i}{2} \mathcal{Y}\,, \\
  \bar M&\equiv \mathcal{J}+\frac{i}{2} \mathcal{Y}\,. 
\end{align}
In order to get a physical sugra theory, torsion constraints have to
be imposed. In our case, the relevant constraints are given by 
\begin{align}
\nonumber \left\{\nabla_\pm, \nabla_\pm\right\}&=0\,,\;\;\; \text{and}\;\;\;  
\left\{\nabla_+, \nabla_-\right\}=-\frac{1}{2}\bar R \bar M\,, \\
\left\{\bar \nabla_\pm,\bar \nabla_\pm\right\}&=0\,,\;\;\;
\text{and}\;\;\;  
\left\{\bar \nabla_+, \bar \nabla_-\right\}=-\frac{1}{2}R M\,, \label{eq::torsionconstraints}\\
\nonumber \left\{\nabla_+, \bar \nabla_-\right\}&=0 \,,\;\;\;\text{and}\;\;\; 
\left\{\bar \nabla_+, \nabla_-\right\}=0 \, ,
\end{align}
where $R$ is the chiral and $\bar R $ the anti-chiral curvature
supermultiplet. These supermultiplets contain in their two $\theta$ component the usual bosonic scalar curvature $\mathcal{R}$ as well as the $U_A(1)$ field strength $\mathcal{F}$ as can be most easily displayed in a Wess-Zumino gauge~\cite{Grisaru:1995dr}:
the components of the supercurvature multiplets can be expressed through the components of the supergravity multiplet, namely the vielbein $e_a{}^m$, the gravitini $\psi_a{}^\alpha$ and the two auxiliary fields $S$ and $\bar S$. These fields are defined by the leading components of the vector covariant derivatives
\begin{align}
     \nonumber \nabla_a| &= \mathbf{D}_a + \psi_a^\alpha \nabla_\alpha| + \psi_a^{\dot \alpha} \nabla_{\dot \alpha}| \\
     &= \mathbf{D}_a + \psi_a^\alpha \partial_\alpha + \psi_a^{\dot \alpha} \partial_{\dot \alpha}\,,\label{eq::nablaa}
\end{align}
where $|$ sets $\theta^+=\theta^-=\bar \theta^+=\bar \theta^-=0$ and 
\begin{align}
    \mathbf{D}_a= e_a + \Omega_a \mathcal{J} + \Sigma_a \mathcal{Y}\,. 
\end{align} 
The leading components of the curvature supermultiplets are given by
\begin{align}
        \left.R\right|=S \,,\;\;\; \left.\bar R\right|=\bar S\,.
\end{align}
 The higher order components can be determined by looking at 
\begin{align}
        \nonumber \left[\nabla_l, \nabla_{\bar l}\right]=&
                                                           -\frac{i}{2} 
\left[(\nabla_- R)\nabla_+ +(\nabla_+ R)\nabla_- 
+ (\bar \nabla_- \bar R)\bar \nabla_+ +(\bar \nabla_+ \bar R)
\bar \nabla_-  \right]\\
    &+ \frac{1}{2}\left[\bar \nabla_-\bar\nabla_+ \bar R 
+\frac{i}{2} \bar R R \right]\bar M
+ \frac{1}{2} \left[ \nabla_- \nabla_+ R + \frac{i}{2} \bar R R\right] M\,.
\end{align}
In the following, we will make use of the notation 
\begin{align}
    \bar \nabla ^2 \equiv \bar \nabla_+\bar \nabla_-\,, \;\;\;\; \nabla^2\equiv \nabla_+\nabla_- \,. 
\end{align}
We can insert (\ref{eq::nablaa}) into the commutator $ \left[\nabla_l, \nabla_{\bar l}\right]$, use
\begin{align}
\left[\nabla_l, \nabla_{\bar l}\right]|= \left[\nabla_l|, \nabla_{\bar l}|\right] + \psi_l^\alpha \nabla_\alpha \nabla_{\bar l}|+\psi_l^{\dot \alpha} \nabla_{\dot \alpha} \nabla_{\bar l}|-\psi_{\bar l}^\alpha \nabla_\alpha \nabla_{ l}|-\psi_{\bar l}^{\dot \alpha} \nabla_{\dot \alpha} \nabla_{ l}|\,,
\end{align}
and read off the other components of $R$ and $\bar R$. The calculation is rather tedious and since we are only interested in a certain classical background solution, we set the gravitini to zero and the only relevant component of the supercurvature is the $\theta^+ \theta^-$ component. This component depends on the $U_A(1)$ field strength $\mathcal{F}$ and the scalar curvature $\mathcal{R}$ and is of the form 
\begin{align}
    \left.\left(\nabla^2 R + \frac{i}{2} \bar R R \right)\right| =-i\left(\mathcal{R} + i \mathcal{F}\right)\,,\label{eq::Rtwotheta}\\
     \left.\left(\bar \nabla^2 \bar R + \frac{i}{2} \bar R R \right) \right| = -i \left(\mathcal{R} - i \mathcal{F} \right)\,,\label{eq::barRtwotheta}
\end{align}
if the gravitini are set to zero. 

Coming back to the torsion constraints (\ref{eq::torsionconstraints}), these are most easily solved in superconformal gauge in terms of a chiral
field $\sigma$ and an anti-chiral field $\bar \sigma$. The solution of the torsion constraints is then given by
\begin{gleichung}
 \nabla_+&=e^{\bar \sigma}\left(D_+ +i (D_+\sigma)\bar M\right)\,,\\
 \nabla_-&=e^{\bar \sigma}\left(D_- -i (D_-\sigma)\bar M\right)\,,\\
 \bar \nabla_+&=e^{\sigma}\left(\bar D_+ +i (\bar D_+\bar \sigma) M\right)\,,\\
 \bar \nabla_-&=e^{\sigma}\left(\bar D_- -i (\bar D_-\bar \sigma) M\right)\,.
\end{gleichung}
The vector derivatives are
\begin{align}
\nonumber \nabla_l=\left\{ \nabla_+, \bar\nabla_+\right\} = e^{\sigma +\bar
  \sigma}&\left[\left(\partial_z+2(D_+\sigma)
\bar D_+ + 2(\bar D_+\bar \sigma)D_+\right)\right.\\
&\hspace*{-1cm}\left.+i\left(\partial_z \bar \sigma +2 (D_+\sigma)(\bar D_+\bar 
\sigma)\right)M+i\left(\partial_z  \sigma +2 (\bar D_+\bar \sigma)(D_+
\sigma)\right)\bar M\right]\,, \\
\nonumber \nabla_{\bar l}=\left\{ \nabla_-,\bar\nabla_-\right\}= e^{\sigma +\bar
  \sigma}&\left[\left(\partial_{\bar z}+2(D_-
\sigma)\bar D_- + 2(\bar D_-\bar \sigma)D_-\right)\right.\\
&\hspace*{-1cm}\left.-i\left(\partial_{\bar z} \bar \sigma +2 (D_-\sigma)(\bar
  D_-\bar 
\sigma)\right)M-i\left(\partial_{\bar z}  \sigma +2 (\bar D_-\bar
  \sigma)(D_-
\sigma)\right)\bar M\right]\,.
\end{align}
The connection and vielbein components can be read off by comparing
our expressions for the covariant derivatives with
(\ref{eq::Nabla1}). One gets for the Lorentz connection
\begin{gleichung}
    \Omega_l \, \, &= i\partial_z(\sigma +\bar \sigma)e^{\sigma +\bar \sigma}\,,\\
    \Omega_{\bar l} \, \, &= -i\partial_{\bar z}(\sigma +\bar \sigma)e^{\sigma +\bar \sigma}\,,\\
    \Omega_+ &= ie^{\bar \sigma}(D_+\sigma)\;\;\;\;\;\; \text{and}\;\;\;\bar \Omega_+=ie^{\sigma}(\bar D_+ \bar \sigma)\,,\\
     \Omega_- &=-ie^{\bar \sigma}(D_-\sigma)\;\;\; \text{and}\;\;\; \bar \Omega_-=-ie^{\sigma}(\bar D_- \bar \sigma)\,.
\end{gleichung} 
The holomorphic part of the vielbein is given by
\begin{align}
E_A{}^\pi=\left(\begin{matrix}\left(1+(D_+\sigma)\theta^++(\bar
D_+\bar \sigma)\bar \theta^+\right)e^{\sigma+\bar
\sigma}&2e^{\sigma+\bar \sigma}(\bar 
D_+\bar \sigma) &2e^{\sigma+\bar \sigma}(D_+\sigma)\\
    \frac{1}{2}e^{\bar \sigma}\bar \theta^+ & e^{\bar \sigma}&0 \\
    \frac{1}{2}e^{\sigma} \theta^+ &0& e^{\sigma}
    \end{matrix}\right)
\end{align}
and the inverse 
\begin{align}
    E_\pi{}^A=\left(\begin{matrix}e^{-(\sigma+\bar \sigma)}&-2(\bar
        D_+\bar 
\sigma)e^{-\bar \sigma}& -2(D_+\sigma)e^{-\sigma}\\
    -\frac{\bar \theta^+}{2}e^{-(\sigma+\bar \sigma)} & e^{-\bar
      \sigma}\left(1+\bar 
\theta^+(\bar D_+\bar \sigma)\right)&e^{- \sigma} \theta^+(\bar D_+\bar \sigma) \\
    -\frac{ \theta^+}{2}e^{-(\sigma+\bar \sigma)}&e^{-\bar \sigma}\bar 
\theta^+(D_+\sigma)& e^{- \sigma}\left(1+ \theta^+( D_+\sigma)\right)
    \end{matrix}\right)\,,
\end{align}
with analogous expressions for the antiholomorphic part.
Finally, the supercurvature is given by
\begin{align}
    R&=4i e^{2 \sigma}(\bar D_+ \bar D_- \sigma)\,, \label{eq::R}\\
    \bar R&= 4i e^{2\bar \sigma}(D_+ D_- \sigma)\,, \label{eq::Rbar}
\end{align}
which are thus respectively a chiral and an anti-chiral superfield.
\section{$\mathbf{{\mathcal N}=(2,2)}$ 
Jackiw-Teitelboim Action
\label{sec::JackiwTeitelboim}}

In the following, we want to consider the ${\mathcal N}=(2,2)$
generalisation to JT 
gravity. First, we consider the action for pure supergravity
supplemented with a Gibbons-Hawking-York term,
\begin{equation}
  S= -\frac{\Phi_0}{16 \pi G_N} \left[\int\limits_\mathcal{M}
    \mathrm{d}^2z \mathrm{d}^2 \theta  \mathcal{E}^{-1} R +
    \int\limits_\mathcal{M} \mathrm{d}^2z \mathrm{d}^2 \bar \theta
    \mathcal{\bar E}^{-1}\bar R + 2  \int\limits_\mathcal{\partial M}
    \mathrm{d}u \mathrm{d} \vartheta K + 2\int\limits_\mathcal{\partial
      M} \mathrm{d}u \mathrm{d} \bar \vartheta \bar
    K\right]\, . \label{eq::GibbonsHawking} 
\end{equation} 
Here, respectively $R$ and $\bar R$ are the chiral and anti-chiral
curvature superfields (\ref{eq::R}), (\ref{eq::Rbar}), $\mathcal{E}$
and $\mathcal{\bar E}$ are the 
chiral and anti-chiral density which are needed to get well-defined
$\text{(anti-)}$ chiral integrals. We comment on the projection to $x$-space at the end of this section. Furthermore, $\Phi_0$ is a constant which can
be interpreted as a constant dilaton, $K$ is the extrinsic curvature
associated to the chiral bulk supercurvature and $\bar K$ is the
anti-chiral extrinsic curvature coming from the anti-chiral
bulk supercurvature. These two extrinsic curvatures can be calculated
from the $\mathcal{N}=(1,1)$ expressions~\cite{Forste:2017kwy} 
\begin{align}
   K&=\frac{T^A \bar D_T n_A}{T^A T_A}\,, \label{eq::chiralK} \\
         \bar K&=\frac{T^A D_T n_A}{T^A T_A}\, , \label{eq::anitchiralK}
 \end{align}
  where $A=l,\, \bar l$.
Furthermore, $T$ is the tangent vector along the boundary,  $n$ the
normal vector satisfying $T^An_A=0$ and $n^An_A=1$ and the derivatives
$\bar D_T$ and $D_T$ are defined as 
 \begin{align}
      D_T n_A&= D n_A + \left(Dz^\xi \Omega_\xi \mathcal{J}\right) n_A\,,\\
       \bar D_T n_A&= \bar D n_A + \left(\bar Dz^\xi \Omega_\xi  \mathcal{J} \right) n_A\,. 
 \end{align}
The supersymmetric generalisations (\ref{eq::chiralK}) and
(\ref{eq::anitchiralK}) of the extrinsic curvature are chosen such
that transformations of the derivatives $D$ and $\bar D$, which
replace the derivative $\partial_u$ in the bosonic extrinsic
curvature, cancel the Berezinian of the $\text{(anti-)}$ chiral superspace
measure (cf.\ \cite{Fu:2016vas}). In the following, it will be
useful to express $K$ and $\bar K$ as a more general boundary
superfield $\mathcal{K}$ in order to couple the extrinsic
curvature to superfields without definite chirality (as e.g.\ the
dilaton at the boundary). We therefore define the overall extrinsic
curvature $\mathcal{K}$ through the condition 
\begin{equation}
   \int\limits_\mathcal{\partial M} \mathrm{d}u \mathrm{d} \vartheta
   \mathrm{d} \bar \vartheta \mathcal{K}\overset{!}{=}
   \int\limits_\mathcal{\partial M} \mathrm{d}u \mathrm{d} \vartheta K
   + \int\limits_\mathcal{\partial M} \mathrm{d}u \mathrm{d} \bar
   \vartheta \bar K\, .
\label{eq:overallK}
\end{equation}
One could also try to directly find an expression for $\mathcal{K}$ by
searching for a generalisation of the derivative $\partial_u$
appearing in the bosonic extrinsic curvature which cancels the
Berezinian of the full $d=1$ superspace measure upon
transformations. However, there is no obvious expression involving the
covariant derivatives $D$ and $\bar D$ and generalising $\partial_u$ that satisfies this condition
since the Berezinian of the full superspace measure equals
one. Therefore, we have to 
take the detour and calculate $\mathcal{K}$ via $K$ and $\bar K$.  

Now, we can use $\mathcal{K}$ to define the supersymmetric
$\mathcal{N}=(2,2)$ generalisation of the JT
action. This action reads  
\begin{align}
    S=& -\frac{1 }{16 \pi G_N} \left[\int\limits_\mathcal{M}
  \mathrm{d}^2 z \mathrm{d}^2 \theta \mathcal{E}^{-1}\Phi ( R-\alpha)+
  \int\limits_\mathcal{M} \mathrm{d}^2 z \mathrm{d}^2 \bar \theta
  \mathcal{E}^{-1}\bar \Phi ( \bar
  R-\alpha)\right. \nonumber \\ &\hspace*{1cm} \left. +2\int\limits_{\partial\mathcal{M}} \mathrm{d}u
  \mathrm{d}\vartheta\mathrm{d}\bar \vartheta  (\Phi_b+\bar
  \Phi_b)\mathcal{K}  \right]\,,\label{eq::bulkaction} 
\end{align}
where $\Phi$ and $\bar \Phi$ are respectively the chiral and
anti-chiral dilaton superfields, which serve as Lagrange multipliers
imposing the constraints $R=\bar R=\alpha$. As we will soon see, the choice $\alpha=-2$ corresponds to an AdS background which we will use from now on. Moreover, $\Phi_b$ and  $\bar \Phi_b$ are the respective
boundary values of the chiral and anti-chiral dilatons. Imposing the
supercurvature constraints yields the effective action for
the boundary degrees of freedom
\begin{equation}
    S_\text{eff}=  -\frac{1 }{8 \pi
  G_N}\int\limits_{\partial\mathcal{M}} \mathrm{d}u
  \mathrm{d}\vartheta\mathrm{d}\bar \vartheta  (\Phi_b+\bar
  \Phi_b)\mathcal{K}\,. \label{eq::effectiveboundaryaction1} 
\end{equation}
One can check that the action above is indeed a supersymmetric generalisation of the bosonic JT action by considering the action in $x$-space, i.e. performing the integrals over the Grassmann variables. To do this, one has to know how to deal with the chiral density.  The procedure to find the expression for the chiral density is explained in detail in \cite{Grisaru:1995dr}: if a Lagrangian $\mathcal{L}$ is considered, the chiral projection has to take the form 
\begin{align}
    \int \mathrm{d}^2 z \mathrm{d}^2 \theta \mathrm{d}^2 \bar \theta E^{-1} \mathcal{L}&= \int \mathrm{d}^2 z \mathrm{d}^2 \theta  \mathcal{E}^{-1} \bar \nabla^2 \mathcal{L}|_{\bar \theta=0}\\
    &= \int \mathrm{d}^2 z e^{-1}\left[\nabla^2+X^+\nabla_++X^- \nabla_- + Y\right]\bar \nabla^2 \mathcal{L}|_{\theta=\bar \theta=0}\,,
\end{align}
where $e=\text{det}(e_a{}^m)$ and the coefficients $X^+$, $X^-$ and $Y$ have to be determined . They can be found from the requirement that the transformation of the full superspace integral to the $x$-space integral should not depend on whether one has a chiral integral or an anti-chiral integral in the intermediate step. As in \cite{Grisaru:1995dr} this condition can be implemented for e.g.\ the kinetic term of a chiral field with $\mathcal{L}=\bar \Phi \Phi$ by choosing $X^+$, $X^-$ and $Y$ s.t.\ the resulting $x$-space integral is symmetric in barred and unbarred quantities. \\
The calculation is tedious and since we are interested in a classical background solution, we set the gravitini to zero again. In that case, we obtain
\begin{align}
    X^+&=X^-=0\,,& Y&=\frac{i}{2} \bar R|=\frac{i}{2} \bar S\,, \\
    \bar X^+&=\bar X^- =0\,,& \bar Y&= \frac{i}{2} R|=\frac{i}{2} S\,,
\end{align}
where $\bar X^+$, $\bar X^-$ and $\bar Y$ are the corresponding quantities for the anti-chiral density projection formula. 

Having the explicit formula for the (anti-)chiral projections, we can now proceed to find the $x$-space action of our particular supergravity setup. Let us for the moment only consider the bulk part of the action. We start with the supersymmetric Einstein-Hilbert action which now reads: 
\begin{align}
    \nonumber S_{EH}=&-\frac{\Phi_0 }{16 \pi G_N} \left[\int\limits_\mathcal{M} \mathrm{d}^2z \mathrm{d}^2 \theta  \mathcal{E}^{-1} R +  \int\limits_\mathcal{M} \mathrm{d}^2z \mathrm{d}^2 \bar \theta   \mathcal{\bar E}^{-1}\bar R \right] \\
    \nonumber =& -\frac{\Phi_0 }{16 \pi G_N}  \left[\int\limits_\mathcal{M} \mathrm{d}^2ze^{-1}\left(\nabla^2 +\frac{i}{2} \bar S\right)R|+\int\limits_\mathcal{M} \mathrm{d}^2ze^{-1}\left(\bar \nabla^2 +\frac{i}{2}  S\right)\bar R|\right] \\
    \nonumber =& -\frac{\Phi_0 }{16 \pi G_N}  \left[\int\limits_\mathcal{M} \mathrm{d}^2ze^{-1}\left(-i (\mathcal{R} + i \mathcal{F})\right)+\int\limits_\mathcal{M} \mathrm{d}^2ze^{-1}\left(-i (\mathcal{R} - i \mathcal{F})\right)\right] \\
    =&  +\frac{i \Phi_0 }{8 \pi G_N} \int\limits_\mathcal{M} \mathrm{d}^2z\, e^{-1}\,\mathcal{R}\,, 
\end{align}
where we made use of (\ref{eq::Rtwotheta}), (\ref{eq::barRtwotheta}). Thus, this part of the action,  together with the extrinsic curvature term just gives the Euler characteristic of $\mathcal{M}$ times an overall prefactor.   

The second part of the bulk action is given by the JT term, which reads (using the chiral projection formula)
\begin{align}
     S_{JT}=&-\frac{1 }{16 \pi G_N} \left[\int\limits_\mathcal{M} \mathrm{d}^2 z \mathrm{d}^2 \theta \mathcal{E}^{-1}\Phi ( R+2)+ \int\limits_\mathcal{M} \mathrm{d}^2 z \mathrm{d}^2 \bar \theta \mathcal{\bar E}^{-1}\bar \Phi ( \bar R+2)\right]\\
    \nonumber =& -\frac{1 }{16 \pi G_N}  \left[\int\limits_\mathcal{M} \mathrm{d}^2ze^{-1}\left(\nabla^2 +\frac{i}{2} \bar S\right)\Phi(R+2)|+\int\limits_\mathcal{M} \mathrm{d}^2ze^{-1}\left(\bar \nabla^2 +\frac{i}{2}  S\right)\bar \Phi (\bar R+2)|\right]\\
    \nonumber  =& \frac{ i }{16 \pi G_N} \int\limits_\mathcal{M} \mathrm{d}^2ze^{-1}\left[\varphi (\mathcal{R}+i \mathcal{F}) + \bar \varphi (\mathcal{R}-i \mathcal{F})-\bar S \varphi - S\bar \varphi +i B(S+2) + i \bar B(\bar S+2) \right]\,, 
\end{align}
where we used (\ref{eq::Rtwotheta}) and (\ref{eq::barRtwotheta}) as well as the component expansion of the dilaton superfield 
\begin{align}
     \Phi=\varphi +\theta^\alpha \lambda_\alpha + \theta^+ \theta^- B \,,\;\;\; \text{and}\;\;\; \bar \Phi=\bar \varphi + \bar \theta^\alpha \bar \lambda_\alpha +\bar \theta^+ \bar \theta^- \bar B\,.
\end{align} 
If we consider the variations of this JT action w.r.t.\ the auxiliary supergravity fields $S$ and $\bar S$, we get the relations
\begin{align}
    B=i\bar \varphi \;\;\;\text{and} \;\;\; \bar B= i \varphi\,. 
\end{align}
Further variations w.r.t.\ the auxiliary dilaton fields $B$ and $\bar B$ yield the bosonic JT action 
\begin{align}
    S_{JT}= \frac{ i }{16 \pi G_N} \int\limits_\mathcal{M} \mathrm{d}^2ze^{-1}\left[\varphi(\mathcal{R}+i\mathcal{F}+2)+\bar \varphi (\mathcal{R}-i\mathcal{F}+2)\right]\,, 
\end{align}
which upon variation w.r.t.\ $\varphi$ and $\bar \varphi$ gives indeed an AdS background with vanishing field strength $\mathcal{F}$. 

Finally, variations w.r.t.\ the vielbein give an energy momentum tensor similar to the bosonic case in \cite{Maldacena:2016upp}. Thus, one possible solution for $\varphi$ and $\bar \varphi$ is given by the dilaton solution found in that reference. This implies in particular that $\varphi=\bar \varphi$.   

\section{Determination of the Superconformal Factor}
A crucial step for calculating the extrinsic curvature is to find an expression for the (anti-) chiral superconformal field $\sigma$ ($\bar \sigma$), which can be done in two different ways: On the one hand, one can consider (\ref{eq::R}), (\ref{eq::Rbar}) and solve for $\sigma$ and $\bar \sigma$ using the constraint $R=\bar R=-2$. On the other hand one can calculate $\sigma$ and $\bar \sigma$ using the Killing spinors of AdS$_2$. Since the final result for the extrinsic curvature and thus the effective boundary action crucially depends on the result for $\sigma$ and $\bar \sigma$, we will present both ways in order to check our findings. 

First, we solve the supercurvature constraints (\ref{eq::R}) and
(\ref{eq::Rbar}) for the superconformal factors $\sigma$ and
$\bar\sigma$. Since $\sigma$ is a chiral superfield it can be written
in  the chiral basis $z_c=z+\frac{1}{2}\theta^+ \bar \theta^+$ 
and  $\bar z_c=\bar z+\frac{1}{2}\theta^- \bar \theta^-$ as
\begin{equation}
    \sigma=\phi(z_c,\bar z_c) + \theta^+ \theta^- w(z,\bar
    z)\,.\label{eq::chiral} 
\end{equation}
Here, $\phi$ and $w$ are functions of the superspace variables which we
will determine later on. Accordingly, the anti-chiral field $\bar
\sigma$ can be written in terms of the anti-chiral basis
$z_{ac}=z-\frac{1}{2}\theta^+ \bar \theta^+$ and $\bar z_{ac}=\bar
z-\frac{1}{2}\theta^- \bar \theta^-$ as  
\begin{equation}
    \bar \sigma = \bar \phi(z_{ac},\bar z_{ac}) + \bar \theta^+ \bar
    \theta^- \bar w(z,\bar z)\,. 
\end{equation}
According to (\ref{eq::Rbar}) $\bar R=-2$ yields
\begin{align}
    w &= \frac{-  i}{2} e^{-2\bar \phi}\,,\\
    0&= 2 w \bar w - \partial_z \partial_{\bar z} \phi \,. \label{eq::Liouville}
\end{align}
The second equation has the form of a Liouville equation. Since we are
interested in an AdS background geometry, we impose  
\begin{equation}
    \bar \phi=\phi= -\frac{1}{2} \log\left(\frac{1}{2 y}\right)\,,
\end{equation}
where $z=t+iy$. With this input, (\ref{eq::Liouville}) can be solved
by setting
\begin{equation}
    \bar w= \frac{- i}{2}e^{-2\phi}\,.
\end{equation}
Note that this implies
$R=-2$, in accordance with the remaining supercurvature
constraint.

After expanding the chiral basis, the superconformal factors will be
given by
\begin{align}
    \sigma=-\frac{1}{2} \log\left(\frac{1}{\beta
  y}\right)+\frac{i}{8y}
\theta^+\bar \theta^+ - \frac{i}{8y} \theta^- \bar \theta^--\frac{i
  }
{4  y} \theta^+ \theta^- -\frac{1}{32y^2} \theta^- \bar \theta^-
  \theta^+ 
\bar \theta^+ \,,\label{eq::sigma}\\
    \bar \sigma=-\frac{1}{2} \log\left(\frac{1}{\beta y}\right)-
\frac{i}{8y}\theta^+\bar \theta^+ + \frac{i}{8y} \theta^- \bar
  \theta^--
\frac{i }{4  y} \bar \theta^+ \bar \theta^-
  -\frac{1}{32y^2} 
\theta^- \bar \theta^- \theta^+ \bar \theta^+\label{eq::barsigma} \,.
\end{align}
We see that $\sigma$ and $\bar \sigma$ are not complex conjugates of each other, thereby making AdS$_2$ a non-unitary background. For our further deliberations we should corroborate the result for the superconformal factors. More precisely, the result of non-unitarity should be confirmed by other means. We opt to perform a short classification of $\mathcal{N}=(2,2)$ supersymmetric backgrounds by calculation of Killing spinors. The following analysis closely follows the steps given in App.\ D of \cite{Gomis:2015yaa}. Since we chose different conventions for our superspace it is useful to re-derive their results for our setup.

Recall, that due to conformal flatness of two-dimensional supergravity, the background geometry is entirely encoded in the conformal factors $e^{-2\sigma}$ and $e^{-2\bar \sigma}$. Thus, the relevant fields that we have to consider in order to determine the Killing spinors $\epsilon$ and $\bar \epsilon$ of supersymmetry variations are just the chiral $\sigma$ field and the anti-chiral $\bar \sigma$ field. To obtain the background geometry, the fermionic components of $\sigma$ and $\bar \sigma$ are set to zero.

 As in \cite{Gomis:2015yaa} the standard restriction for supersymmetry can be written as
\begin{align}
    \partial_z \epsilon^-=\partial_z \bar \epsilon^-=0\,,\;\;\;\text{and}\;\;\; \partial_{\bar z} \epsilon^+=\partial_{\bar z} \bar \epsilon^+=0\,. 
\end{align}
Further restrictions on the Killing spinors come from the requirement, that the fermionic components of the conformal factors $e^{-2 \sigma}$ and $e^{-2 \bar \sigma}$ remain zero under local supersymmetry transformations. In the following, we will work in the chiral basis as introduced above (\ref{eq::chiral}). With our conventions these conditions can be written as
\begin{align}
    \partial_{z_c} \left(\bar \epsilon^+ e^{-2\phi} \right) -2 \epsilon^- w e^{-2\phi}&=0\,,\;\;\; \partial_{\bar {z_c}} \left(\bar \epsilon^- e^{-2\phi} \right) +2 \epsilon^+ w e^{-2\phi}=0\,,\label{eq::Killing2}\\
    \partial_{z_{ac}} \left( \epsilon^+ e^{-2 \bar \phi} \right) -2 \bar \epsilon^-  \bar w e^{-2\bar \phi}&=0\,,\;\;\; \partial_{\bar z_{ac}} \left( \epsilon^- e^{-2\bar \phi} \right)+2 \bar \epsilon^+ \bar w e^{-2\bar \phi}=0\,,\label{eq::Killing3}
\end{align}
where we inserted $\sigma=\phi+\theta^+\theta^- w$ and $\bar \sigma=\bar \phi+\bar \theta^+\bar \theta^- \bar w$. 
The classification of backgrounds preserving different numbers of supercharges can now be carried out along the lines of \cite{Gomis:2015yaa}.

For a background preserving one supercharge with a particular $U_A(1)$ charge, we can e.g.\ choose the Killing spinor ($\epsilon_1^+$, $\bar \epsilon_1^-$) to be non-zero with the other Killing spinor components zero. Solving (\ref{eq::Killing2}), (\ref{eq::Killing3}) algebraically, we get
\begin{align}
    w&=\frac{1}{2}\frac{\bar \epsilon_1^-}{\epsilon_1 ^+} \partial_{\bar z_c}\left(2\phi -\log \bar \epsilon_1^- \right) \,,\label{eq::w1}\\
    \bar w&=-\frac{1}{2}\frac{ \epsilon_1^+}{\bar \epsilon_1 ^-} \partial_{ z_{ac}}\left(2\bar \phi -\log \epsilon_1^+ \right)\label{eq::barw1}\,.
\end{align}
If the background should also preserve a second supercharge of the opposite $U_A(1)$ charge, there should also exist a second non-zero Killing spinor  ($ \epsilon_2 ^-$, $\bar \epsilon_2 ^+$). Solving (\ref{eq::Killing2}), (\ref{eq::Killing3}) with this Killing spinor yields results similar to (\ref{eq::w1}), (\ref{eq::barw1}). Consistency of the two solutions requires 
\begin{align}
    \left(\bar \epsilon^-_1 \epsilon_2^-\partial_{\bar z_{ac}}+\bar \epsilon^+_2\epsilon_1^+\partial_{z_{ac}}\right)\left(2\bar \phi +\log \epsilon_1^+ \epsilon_2^- \right)=0\,,\label{eq::consistency}
\end{align}
with a similar expression for $z_c$ and $\phi$. Thus, $\bar \phi$ is invariant under the vector $v=\bar \epsilon^-_1 \epsilon_2^-\partial_{\bar z_{ac}}+\bar \epsilon^+_2\epsilon_1^+\partial_{z_{ac}}$ up to a superconformal transformation. 

As shown in e.g.\ \cite{Closset:2014pda}, maximally four supercharges are preserved if and only if the background space is maximally symmetric and the $U_A(1)$ gauge field has zero field strength.
Thus, with (\ref{eq::w1}), (\ref{eq::barw1}) the $w$ and $\bar w$ fields can be expressed in terms of the bosonic conformal factor $\phi$. Since we are interested in an AdS background, we know that $\phi=-\frac{1}{2}\log\left(\frac{1}{2y_c}\right)$ and $\bar \phi=-\frac{1}{2}\log\left(\frac{1}{2y_{ac}}\right)$. In that case indeed a set of four Killing spinors satisfying (\ref{eq::consistency}) can be found. These are
\begin{align}
    \zeta\equiv\left(\begin{matrix}\epsilon^+ \\ \bar \epsilon^-\end{matrix}\right)=\frac{1}{2} \left(\begin{matrix}-1\\ 1\end{matrix}\right)\,,\;\;\; \bar \zeta \equiv \left(\begin{matrix} \epsilon^- \\ \bar  \epsilon^+\end{matrix}\right)= \frac{1}{2} \left(\begin{matrix}-1\\ 1\end{matrix}\right)\,, \label{eq::Killingspinors1}
\end{align}
and 
\begin{align}
        \eta\equiv\left(\begin{matrix}\epsilon^+ \\ \bar \epsilon^-\end{matrix}\right)= \frac{1}{2}\left(\begin{matrix}- z\\ \bar z\end{matrix}\right)\,,\;\;\; \bar \eta \equiv \left(\begin{matrix}\epsilon^- \\ \bar  \epsilon^+\end{matrix}\right)=\frac{1}{2} \left(\begin{matrix}- \bar z\\  z\end{matrix}\right)\,, \label{eq::Killingspinors2}
\end{align}
where we used a bar to distinguish Killing spinors with opposite $U_A(1)$ charge. Indeed, with these four Killing spinors, one can now calculate the three Killing vectors $\zeta \gamma^\mu \bar \zeta \partial_\mu $, $\eta \gamma^\mu \bar \zeta \partial_\mu $ and $\eta \gamma^\mu \bar \eta \partial_\mu $ to be 
\begin{align}
 L_{-1}&\equiv\zeta\gamma^\mu \bar \zeta \partial_\mu = -\frac{1}{2}\left(\partial_z+\partial_{\bar z}\right)\,,\\
L_{0}&\equiv \eta \gamma^\mu \bar \zeta \partial_\mu = -\frac{1}{2}\left( z\partial_z+\bar z\partial_{\bar z}\right)\,,\\
    L_1&\equiv\eta\gamma^\mu \bar \eta \partial_\mu = -\frac{1}{2}\left(z^2\partial_z+\bar z^2\partial_{\bar z}\right)\,,
\end{align}
which are precisely the Killing vectors of AdS$_2$. At the boundary ($y\rightarrow 0$) they correctly reduce to the global conformal transformations 
\begin{align}
    L_{0} \rightarrow - t\partial_t\,,\;\;\; L_{-1}\rightarrow - \partial_t\,,\;\;\; L_1\rightarrow - t^2\partial_t\,. 
\end{align}
Now one can take any  Killing spinor out of (\ref{eq::Killingspinors1}), (\ref{eq::Killingspinors2}) to calculate $w$ and $\bar w$. The results for $\sigma$ and $\bar \sigma$ are 
\begin{align}
    \sigma&=-\frac{1}{2}\log\left(\frac{1}{2y_c}\right) - \frac{i}{4y_c}\theta^+\theta^-\,,\\
    \bar \sigma&=-\frac{1}{2}\log\left(\frac{1}{2y_{ac}}\right) - \frac{i}{4y_{ac}}\bar \theta^+\bar \theta^-\,.
\end{align}

These results obtained by considering the Killing spinors perfectly coincide with our solution for $\sigma$ and $\bar \sigma$ (\ref{eq::sigma}), (\ref{eq::barsigma}) as obtained from the requirement $\bar R=R=-2$. In particular, this shows again that our background is non-unitary since $\sigma$ and $\bar \sigma$ are not complex conjugates of each other.
\section[Effective Action: Appearance of the 
Super-Schwarzian]{Effective Action:\\ Appearance of the
Super-Schwarzian}  
Only boundary curves of constant arc length are considered in the calculation of the effective action \cite{Maldacena:2016upp,Forste:2017kwy},
\begin{equation}
    \frac{\mathrm{d}u^2+\bar \vartheta \mathrm{d}\vartheta
      \mathrm{d}u+\vartheta \mathrm{d}\bar \vartheta
      \mathrm{d}u+\frac{1}{2}\vartheta\bar\vartheta \mathrm{d}
      \vartheta\mathrm{d}\bar \vartheta}{4\epsilon^2}=
    \left.\left(\mathrm{d}z^{\xi}E_\xi^l \mathrm{d}z^{\pi}E_\pi^{\bar
          l}\right)\right|_{\text{pull-back}} \,.\label{eq::pullback} 
\end{equation}
This results in the constraints
\begin{align}
    Dz&= \frac{1}{2}\left(\bar \theta^+(D\theta^+) + \theta^+ ( D \bar
        \theta^+)\right)\,, \;\;\;\bar D z= \frac{1}{2}\left(\theta^+
        (\bar D \bar \theta^+)+\bar \theta^+(\bar D\theta^+)\right)
        \,\label{eq::Dz},\\ 
    D\bar z&= \frac{1}{2}\left( \theta^-(D\bar\theta^-)+\bar \theta^-
             ( D \theta^-)\right)\,,\;\;\; \bar D \bar z=
             \frac{1}{2}\left(\bar \theta^- (\bar D
             \theta^-)+\theta^-(D\bar\theta^-)\right)\, , 
             \label{eq::Dbarz} 
\end{align}
where we defined one dimensional supercovariant derivatives as in 
\cite{Fu:2016vas} (up to factors of one half due to differences in conventions), 
\begin{equation}
D = \frac{\partial}{\partial \vartheta}
+\frac{\bar\vartheta}{2}\frac{\partial}{\partial u} \,\,\, ,\,\,\, \bar D =
\frac{\partial}{\partial \bar\vartheta} 
+\frac{\vartheta}{2}\frac{\partial}{\partial u}  
\end{equation}
In addition to (\ref{eq::Dz}) and (\ref{eq::Dbarz}), we impose the
following chirality conditions 
\begin{equation}
    D \theta^- = D \bar \theta^+ = \bar D \theta ^+ = \bar D \bar
    \theta^- =0\,. \label{eq::dz0} 
\end{equation}

Equations (\ref{eq::Dz}), (\ref{eq::Dbarz}) and (\ref{eq::dz0}) are equivalent to the $\mathcal{N}=2$ superconformal transformations of \cite{Fu:2016vas}. The bulk variables correspond to super-reparametrisations of the boundary.
Furthermore, the conformal factor has to satisfy
\begin{equation}
    e^{2(\sigma +\bar \sigma)}= 4\epsilon^2 \left[(D \theta^+)(D \bar
  \theta^-)(\bar D \theta ^-)(\bar D \bar \theta^+) \right] \,. 
\end{equation}
Together with (\ref{eq::sigma}), (\ref{eq::barsigma}) this leads to
\begin{equation}
    y=\text{Im} \, z =\epsilon \left[(D \theta^+)(D
      \bar \theta^-)(\bar D \theta ^-)(\bar D \bar \theta^+)
    \right]^{1/2}+ \frac{i }{4 }\left(\theta^+\theta^-+
      \bar \theta^+ \bar\theta^-\right)\,. 
\end{equation}
Now, we can calculate the chiral and anti-chiral part of the extrinsic
curvature given in (\ref{eq::chiralK}) and
(\ref{eq::anitchiralK}).  
The tangent vector for the boundary can be evaluated using 
\begin{align}
    T^l=(\partial_u z^\xi)E_\xi^l\,, \\
    T^{\bar l}=(\partial_u z^\xi)E_\xi^{\bar l}\, ,  
\end{align}
leading to
\begin{align}
    T^l=e^{-(\sigma+\bar \sigma)}\left[(D \theta^+)(\bar D \bar
  \theta^+)\right]\,,\\ 
    T^{\bar l}=e^{-(\sigma+\bar \sigma)}\left[(D \bar  \theta^-)(\bar
  D  \theta^-)\right]\,, 
\end{align}
and
\begin{align}
    n_l=-\frac{i}{2}\left(\frac{(D \bar  \theta^-)(\bar D
  \theta^-)}{(D \theta^+)(\bar D \bar \theta^+)}\right)^{1/2}\, ,\\ 
    n_{\bar l}=+\frac{i}{2}\left(\frac{(D \theta^+)(\bar D \bar
  \theta^+)}{(D \bar  \theta^-)(\bar D  \theta^-)}\right)^{1/2} .
\end{align}
Hence, the contribution to the anti-chiral extrinsic supercurvature $\bar K$
which does not include the connection is given by
\begin{align}
    \frac{T^l D n_l+T^{\bar l} D n_{\bar l} }{T^2}= i \epsilon
  \left[\frac{\left(\bar \theta^+\right)^\prime}{\left(\bar D \bar \theta
  ^+\right)}-\frac{\left( \theta^-\right)^\prime}{\left(\bar D \theta
  ^-\right)}\right]\,, 
\end{align}
with a similar expression for the chiral extrinsic supercurvature $
K$. Here, the prime indicates derivatives with respect to $u$. For the
part of $\bar K$ containing the connection, we first observe that the
Lorentz generators applied to $n_l$ and $n_{\bar l}$ give 
\begin{align}
    \left[\mathcal{J} , n_{\bar l}\right]=i n_{\bar l}\,,\;\;\;\left[\mathcal{J} , n_{
  l}\right]=-i n_{ l}\,. 
\end{align}
Thus, the contribution to $\bar K$ containing the connection part is
given by
\begin{align}
    \frac{1}{T^2}&T^A(Dz^\xi\Omega_\xi)J n_A=-2 \epsilon(Dz^\xi\Omega_\xi)\\
    &=\frac{1}{ 2\left[(D \theta^+)(D \bar \theta^-)(\bar D \theta
      ^-)(\bar D \bar \theta^+) \right]^{1/2}}\left[(D \theta^+)(\bar
      \theta^+-\theta^-)+(D \bar \theta^-)
      (\theta^--\bar \theta^+)\right]\,.  
\end{align}
Having the general expression for the extrinsic curvature, we want to
make contact to the boundary theory. The $\mathcal{N}=(2,2)$
supersymmetry of the bulk reduces to $\mathcal{N}=(1,1)$ supersymmetry
on the boundary.
We therefore
need an expression for the bulk variables at the boundary in terms of
the boundary degrees of freedom. To zeroth order the solution to
(\ref{eq::pullback}) after imposing (\ref{eq::Dz}), 
(\ref{eq::Dbarz}) reads 
\begin{equation}
    \theta^+=\bar \theta^-=\xi\,,\;\;\;  \theta^-=\bar \theta^+=\bar
  \xi\,,\,\;\;\;\text{Im}z=\epsilon(D\xi) (\bar D \bar
  \xi)\,. 
\end{equation}
In that case, (\ref{eq::dz0}) reduces to 
\begin{equation}
    D\bar \xi=\bar D\xi=0\,.
\label{eq:chircon}
\end{equation}
We also need the corrections in $\epsilon$ to these solutions. We
choose the ansatz 
\begin{align}
    \theta^+= \xi -i\epsilon  \rho \,,\;\;\; \bar \theta^+=\bar \xi
  +i\epsilon  \bar \rho \,,\label{eq::thetaplus}\\ 
        \theta^-= \bar \xi -i\epsilon \bar \rho \,,\;\;\; \bar
  \theta^-=\xi +i\epsilon \rho \,.\label{eq::thetaminus} 
\end{align}
With this ansatz, (\ref{eq::Dz}), (\ref{eq::Dbarz}) can be solved by
\begin{align}
    \rho=-\xi^\prime\;\;\;\text{and} \;\;\; \bar
  \rho=\bar \xi^\prime\,.\label{eq::thetasolution} 
\end{align}
Thus, the boundary
solution of $\theta^+, \theta^-, \bar \theta^+, \bar \theta^-$ has the
form of a Taylor expansion of e.g.\ $\theta^+(u+i\epsilon, \vartheta,
\bar \vartheta)$ around $\epsilon=0$, with $\theta^+(u, \vartheta,
\bar \vartheta)=\xi$ and similarly for the other Grassmann
coordinates.

With this solution, the extrinsic curvature is given by
\begin{align}
    \bar K&=-2\epsilon^2\left[\frac{\bar \xi^{\prime\prime}}{\bar D
            \xi}
-\frac{\bar
            \xi^\prime (\bar D \bar \xi^\prime)}{(\bar D\bar \xi)^2}+
            \left(\frac{(D\xi^\prime)\bar \xi^\prime}{(D\xi)(\bar D\bar
            \xi)}\right)\right]\,, \label{eq:barK}\\ 
      K&=-2\epsilon^2\left[\frac{ \xi^{\prime\prime}}{ D \xi}-\frac{
         \xi^\prime ( D
         \xi^\prime)}{( D \xi)^2}+ \left(\frac{(\bar D\bar \xi^\prime)
         \xi^\prime}{(D\xi)(\bar D\bar \xi)}\right)\right]\,. 
\label{eq:K}
\end{align}
The next step is to find the overall extrinsic curvature ${\mathcal
  K}$ as defined in (\ref{eq:overallK}). The following identity proves
useful
\begin{equation}
\hspace*{-2mm}
\int \mathrm{d}u \mathrm{d}\vartheta \mathrm{d}\bar\vartheta \, f\!\left( u,
  \vartheta,\bar\vartheta\right) = \int 
  \mathrm{d}u_c \mathrm{d}\vartheta \left(\bar D
  f\right)\!\left(\!u+\frac{\vartheta\bar\vartheta}{2},
  \vartheta\! \right)= - \!\int \mathrm{d}u_{ac}  
\mathrm{d}\bar\vartheta \left( D f\right)\!\left( \!u -\frac{\vartheta\bar\vartheta}{2},
\bar\vartheta\!\right) ,
\label{eq:usfid}
\end{equation}
where $\mathrm{d}u \simeq \mathrm{d}u_{c/ac}=\mathrm{d}\left( u \pm
  \frac{1}{2}\vartheta\bar\vartheta\right)$ denote chiral and anti-chiral
measures. To check this formula it is best to expand the superfield
$f$ into components
\begin{equation}
f\left( u , \vartheta, \bar\vartheta\right) = g\left( u\right) +
\vartheta \bar\zeta\left(u\right)+\bar\vartheta\zeta\left(u\right) +
\vartheta\bar\vartheta h\left( 
  u\right) .
\end{equation}
Then one gets
\begin{equation}
\int \mathrm{d}u \mathrm{d}\vartheta d\bar\vartheta \, f\left( u,
  \vartheta,\bar\vartheta\right) = -\int \mathrm{d}u h .
\label{eq:total}
\end{equation}
whereas for the second integral one gets
\begin{equation}
\int \mathrm{d}u_c \mathrm{d}\vartheta \left(\bar D
  f\right)\!\left(u+\frac{\vartheta\bar\vartheta}{2},\theta \right)
= \int \mathrm{d}u_c \mathrm{d}\vartheta \left[ \zeta\left( u_c\right) + \vartheta
  \left(\frac{g^\prime\left( u_c\right)}{2}
  - h\left( u_c\right)\right)\right] = -\!\int
  \mathrm{d}u h .
\end{equation}
In the second step the first term has been killed by the $\mathrm{d}\vartheta$
integral and the second term gave rise to a $\mathrm{d}u_c$ integral over a
derivative, leaving only the third contribution in agreement with
(\ref{eq:total}). The other equality in (\ref{eq:usfid}) can be seen
analogously.

Note that $\bar K$ and $K$ in (\ref{eq:barK}), (\ref{eq:K}) can be 
expressed as derivatives,
\begin{align}
    \bar K= -2 \epsilon^2  D\left( \frac{(\bar D \bar\xi^\prime)}{\bar D
  \bar \xi} - \frac{\xi^\prime\bar \xi^\prime}{(D\xi)(\bar D \bar \xi)} \right)
  \,,\\ 
     K = -2 \epsilon^2 \bar D \left(\frac{( D \xi^\prime)}{ D  \xi}+
  \frac{\xi^\prime\bar \xi^\prime}{(D\xi)(\bar D \bar \xi)}\right) \,.  
\end{align}
This result can now be plugged into the Gibbons-Hawking-York term in
(\ref{eq::GibbonsHawking}):  
\begin{align}
  \nonumber  S_{GH}=&  -\frac{1 }{8 \pi G_N\epsilon^2}\int\limits_{\partial
                      \mathcal{M}} \mathrm{d} u \mathrm{d} \bar
                      \vartheta \left[-2 \epsilon^2 D\left(
                      \frac{(\bar D \bar\xi^\prime)}{\bar D \bar \xi} -
                      \frac{\xi^\prime\bar \xi^\prime}{(D\xi)(\bar D \bar \xi)}
                      \right)\right] \\ 
  & -\frac{1 }{8 \pi G_N\epsilon^2} \int\limits_{\partial \mathcal{M}}
    \mathrm{d} u \mathrm{d}  \vartheta \left[ -2 \epsilon^2 \bar D
    \left(\frac{( D \xi^\prime)}{ D  \xi}+ \frac{\xi^\prime\bar
    \xi^\prime}{(D\xi)(\bar 
    D \bar \xi)}\right)\right]\, ,  
\end{align}
where the $1/\epsilon^2$ factor arises due to the $\epsilon$ factor in the
flat space vielbein (\ref{eq::pullback}) \cite{Forste:2017kwy}.  
With (\ref{eq:usfid}) each integral in the above expression can be
expressed as an integral over the total superspace. Comparing with
(\ref{eq:overallK}) yields 
\begin{equation}
{\mathcal K} = 2 \epsilon^2 \text{Schw}\left( t,\xi, \bar \xi ;
  u, \vartheta, \bar \vartheta \right)\,. 
\end{equation}
where
\begin{equation}
\text{Schw}\left( t,\xi, \bar \xi ;
  u, \vartheta, \bar \vartheta \right) =  \frac{(\bar D \bar\xi^\prime)}
{\bar D \bar \xi}- \frac{( D \xi^\prime)}{ D  \xi}-2
\frac{\xi^\prime
\bar \xi^\prime}{(D\xi)(\bar D \bar \xi)} 
\end{equation}
denotes the super-Schwarzian. 
With this expression for the extrinsic curvature, we get for the
effective boundary action in (\ref{eq::effectiveboundaryaction1})  
\begin{align}
    S_\text{eff}=-\frac{1}{4 \pi G_N}
  \int\limits_{\partial\mathcal{M}} \mathrm{d}u
  \mathrm{d}\vartheta\mathrm{d}\bar \vartheta  (\Phi_b+\bar \Phi_b)
  \text{Schw}(t,\xi, \bar \xi ; u, \vartheta, \bar \vartheta
  )\,, 
\end{align}
which can be further simplified by noting that only the leading components
$\varphi$ of the dilaton supermultiplets contribute at the boundary,
since  for the two $\theta$ components of the dilaton, at the boundary
we have at zeroth order in $\epsilon$ the relation 
\begin{equation}
    \left(\Phi+\bar\Phi\right)_b\supset i\left(\varphi \theta^+
  \theta^- + \varphi \bar \theta^+ \bar \theta^-\right)_b= i \varphi_b
  (\xi \bar \xi + \bar \xi \xi)=0\,, 
\label{eq:effect}
\end{equation}
where $\varphi_b$ is the value of the leading component at the
boundary and we used that the chiral and the anti-chiral dilaton
superfields have the same leading component on-shell
\begin{align}
    S_\text{eff}=-\frac{1 }{2 \pi G_N}
  \int\limits_{\partial\mathcal{M}} \mathrm{d}u
  \mathrm{d}\vartheta\mathrm{d}\bar \vartheta \, \varphi_b\,
  \text{Schw}(t,\xi, \bar \xi ; u, \vartheta, \bar \vartheta
  )\,. 
\label{eq:effective}
\end{align}

We close this section by briefly commenting on the physical
interpretation of the action considered here (cf. e.g.\ \cite{Fu:2016vas}, \cite{Sarosi:2017ykf} for the bosonic case). The entire action reads 
\begin{align}
   \nonumber  S=& -\frac{\Phi_0}{16 \pi G_N}
                  \left[\int\limits_\mathcal{M} \mathrm{d}^2z
                  \mathrm{d}^2 \theta  \mathcal{E}^{-1} R +
                  \int\limits_\mathcal{M} \mathrm{d}^2z \mathrm{d}^2
                  \bar \theta   \mathcal{\bar E}^{-1}\bar R + 2
                  \int\limits_\mathcal{\partial M} \mathrm{d}u
                  \mathrm{d} \vartheta \mathrm{d} \bar \vartheta
                  \mathcal{K}\right]\\ 
   &-\frac{\Phi_0}{16 \pi G_N}\left[\int\limits_\mathcal{M}
     \mathrm{d}^2 z \mathrm{d}^2 \theta \mathcal{E}^{-1}\Phi (
     R+2) 
  + \int\limits_\mathcal{M} \mathrm{d}^2 z \mathrm{d}^2 \bar \theta
     \mathcal{\bar E}^{-1}\bar \Phi ( \bar
     R+2)\right. \label{eq::fullaction}\\ &\left. \hspace{2.5cm}+2\int\limits_{\partial\mathcal{M}} \mathrm{d}u
     \mathrm{d}\vartheta \mathrm{d}\bar \vartheta (\Phi_b+\bar \Phi_b)
     \mathcal{K} \right]\,.\nonumber 
\end{align}
The manifold  $\mathcal{M}$ is obtained by cutting out a line
given by coordinates $t(u, \vartheta, \bar \vartheta)$, $y(u,
\vartheta, \bar \vartheta)$, $\theta^+(u, \vartheta, \bar
\vartheta)\,, \; \theta^-(u, \vartheta, \bar \vartheta)\,,\; \bar
\theta^+(u, \vartheta, \bar \vartheta)$ and $\bar  \theta^-(u,
\vartheta, \bar \vartheta)$ corresponding to the boundary $\partial \mathcal{M}$. The parameters $u\,,\; \vartheta\,,\;
\bar \vartheta $ are the coordinates of the one-dimensional
$\mathcal{N}=(1,1)$ superspace at this boundary. Different
boundaries are related via superreparametrisations of the
one-dimensional superspace. We saw earlier that the first line of (\ref{eq::fullaction}) is just topological and corresponds to the Euler characteristic of $\mathcal{M}$ which is not altered by superreparametrisations. Thus, we have a large  symmetry
group consisting of all superreparametrisations 
that do not violate the chirality constraints on the superfields in
(\ref{eq::Dz})-(\ref{eq::dz0}). We derived in
(\ref{eq::thetaplus})-(\ref{eq::thetasolution}), under the
assumption of the chirality constraints, that our reparametrisations
satisfy  
\begin{align}
    \theta^+&= \xi (u, \vartheta, \bar \vartheta) + i\epsilon \xi'(u,
              \vartheta, \bar \vartheta) \,,\;\;\;  \bar \theta^+=
              \bar \xi (u, \vartheta, \bar \vartheta) + i\epsilon \bar
              \xi'(u, \vartheta, \bar \vartheta)\,,\\ 
    \theta^-&= \bar \xi (u, \vartheta, \bar \vartheta) - i\epsilon
              \bar  \xi'(u, \vartheta, \bar \vartheta) \,,\;\;\;  \bar
              \theta^-= \xi (u, \vartheta, \bar \vartheta) -
              i\epsilon \xi'(u, \vartheta, \bar \vartheta)\,,\\ 
    y&= \epsilon (D \xi )(\bar D \bar \xi) - \frac{\epsilon}{2}(\xi
       \bar \xi' + \bar \xi \xi ') \,.  
\end{align}
Since superreparametrisations of the one-dimensional superspace in general
map a given boundary to a
completely different one, we have a spontaneous breaking of the entire
superreparametrisation symmetry to the subgroup $SU(1,1|1)$ of
global reparametrisations that leave the boundaries invariant
\cite{Fu:2016vas}. As in the bosonic case 
the other reparametrisations can be interpreted as 
Goldstone modes.
  
The part of the action in (\ref{eq::fullaction}) involving the
dilaton now explicitly breaks this symmetry since its boundary term
gives a non-zero action for the superreparametrisations, namely the
Schwarzian action in (\ref{eq:effective}) which vanishes
only for $SU(1,1|1)$ reparametrisations~\cite{Fu:2016vas}. 

\section{Consistency Check}

Here, we perform the following consistency check. The
Gibbons-Hawking-York term should ensure that Dirichlet conditions on
field variations should not lead to further boundary conditions on 2d
fields. Hence, any solution to the bulk equations should be viable. In
particular it can be seen that plugging a bulk solution of the dilaton
into the effective action (\ref{eq:effect}) and taking variations
with respect to superreparametrisations yields zero. For cases with
less supersymmetry this has been done in
\cite{Maldacena:2016upp,Forste:2017kwy}. In particular, in
\cite{Forste:2017kwy} superreparametrisations have been expressed in
terms of unconstrained bosonic and fermionic degrees of freedom with
respect to which the variation has been considered. Already there,
this procedure turned out to give rather lengthy expressions. In the
${\mathcal N}=2$ case the complexity of this calculation grows
further \cite{joshua2018}. Fortunately, there is a shortcut which
could have been used also in \cite{Forste:2017kwy}. The underlying
trick can be found e.g.\ in chapter 4 of
\cite{Blumenhagen:2013fgp}. Using the anomalous chain rule, the
variation of the Schwarzian can be linearised. For ${\mathcal N}= 2$
the details are as follows,
\begin{align}
\delta \text{Schw}\left( t, \xi, \bar\xi ; u, \vartheta,
  \bar\vartheta\right) & \equiv 
\text{Schw}\left( t +\delta t,
  \xi+\delta\xi, \bar\xi +\delta\bar\xi ; u, \vartheta,
  \bar\vartheta\right) - \text{Schw}\left( t, \xi, \bar\xi ; u, \vartheta,
  \bar\vartheta\right) \nonumber \\ & 
= D\xi\bar D\bar\xi \text{Schw}\left( t +\delta t,
  \xi+\delta\xi, \bar\xi +\delta\bar\xi ; t, \xi,
  \bar\xi\right)\nonumber\\& 
=  D\xi\bar D\bar\xi\left(  \partial_t  \bar D _\xi
\delta\bar\xi - \partial_t D_\xi \delta\xi\right) .
\end{align}
The next steps are to replace bosonic derivatives by anti-commutators
of supercovariant derivatives and use 
$$ D_\xi = \left( D\xi\right)^{-1} D \,\,\, ,\,\,\, \bar D_\xi =
\left( \bar D\bar \xi\right)^{-1} \bar D .$$
For the variation of (\ref{eq:effective}) one gets
\begin{align}
\delta S \sim & \nonumber \\
&\hspace*{-0.5cm}\int \mathrm{d}u \mathrm{d}\vartheta \mathrm{d}\bar\vartheta \left\{ \delta \bar \xi
  \bar D \left( \frac{1}{\bar D\bar \xi}D\left( \frac{1}{D\xi}\bar
      D\left( \varphi_b D\xi\right)\right)\right) - 
 \delta \xi
  D \left( \frac{1}{ D\xi}\bar D\left( \frac{1}{\bar D\bar \xi}
      D\left( \varphi_b \bar D\bar \xi\right)\right)\right)\right\}
\label{eq:efvar}
\end{align}
From the discussion in Sec.~\ref{sec::JackiwTeitelboim} one can see
that $\varphi_b$ is given by the ${\mathcal N}=0$ solution
\cite{Maldacena:2016upp}, multiplied with appropriate powers of
$\epsilon$, evaluated at the boundary
\begin{equation}
\varphi_b = \frac{\alpha + \beta t  + \gamma t^2}{D\xi \bar
  D\bar \xi} ,
\end{equation}
with
\begin{equation} 
2 Dt =  \bar \xi D\xi \,\,\, ,\,\,\, 2\bar D t = 
\xi \bar D\bar \xi\,.
\label{eq:surep}
\end{equation}
Plugging this into (\ref{eq:efvar}) and using the chirality
conditions (\ref{eq:chircon}) yields
\begin{equation}
\delta S_{\text{eff}} \sim \gamma\int \mathrm{d}u \mathrm{d}\vartheta \mathrm{d}\bar \vartheta
\left( \xi \delta \bar \xi - \bar \xi \delta \xi\right) .  
\end{equation}
With (\ref{eq:usfid}) and the variation of  (\ref{eq:surep}) this can
be brought into the form
\begin{equation}
\delta S_{\text{eff}} \sim \int \mathrm{d}u \mathrm{d}\vartheta \bar D \delta t =
\int \mathrm{d}u \delta t^\prime = 0\,.
\end{equation}

\section{Summary and Conclusions}
In the present paper, we argued that the gravitational part of the holographic dual
to the ${\mathcal N}=2$ supersymmetric SYK model is given by an
${\mathcal N}= \left( 2,2\right)$ supersymmetric JT action. We elaborated on the construction of this supersymmetric extension including also a 
Gibbons-Hawking-York boundary term. 
The main part of the
calculation was done in superconformal gauge. 
The superconformal factors can be determined in two ways giving the same result. First, one can solve
the constraints of constant supercurvatures. On
the other hand imposing the existence of four unbroken supersymmetries together with an AdS$_2$ metric yields the same superconformal factors.
Symmetry breaking patterns due to a UV regulator 
curve match those of the SYK model. Further, we showed that the effective 
Lagrangian of those curves is given by the super-Schwarzian in agreement with the 
result \cite{Fu:2016vas} for SYK. 
The chirality of the SYK model emerges from the chirality of the two separate extrinsic curvature fields in our gravitational setup. As a consistency check
we plugged a known dilaton solution into the
effective boundary action. Its variation with respect to super-reparametrisations vanishes. 

It would be interesting to see whether there are corrections towards deviation 
from  JT. For ${\mathcal N}=0$ such corrections have been proposed in \cite{Kitaev:2017awl}. Also the reconstruction of a more complete holographic dual
along the lines of \cite{Gross:2017hcz,Gross:2017aos} should be extended to 
supersymmetric models. A further subject of future research will be to study if JT supergravity admits more 
general dilaton bulk solutions than the one considered here.  Less supercharges than in the
constant dilaton case will be preserved.

Supersymmetric JT gravity  might also be 
considered in the context of $d=4$ black hole 
physics with extended supersymmetry. In particular, the qualitative difference between $\mathcal{N}=1$ and $\mathcal{N}=2$ SYK, the non-perturbative SUSY breaking of the former \cite{Fu:2016vas}, 
should be relevant in this setting. 

\section*{Acknowledgements}  
We thank Iris Golla  and Hans Jockers for useful discussions.
This work was supported by the SFB-Transregio TR33 ``The Dark
Universe'' (Deutsche Forschungsgemeinschaft) and by ``Bonn-Cologne
Graduate School for Physics and Astronomy'' (BCGS).

\end{document}